\documentstyle[12pt,aasms4]{article}

\def\yrm1m{{\rm yr}^{-1}}
\def\yrm1{yr${}^{-1}$}

\def\msun{$M_{\sun}$~}

\def\msunend{$M_{\sun}$}

\begin{document}

\title{ON THE VERY LONG TERM EVOLUTIONARY BEHAVIOR OF HYDROGEN-ACCRETING
LOW MASS CO WHITE DWARFS}

\author{Luciano Piersanti \altaffilmark{1,2}, Santi Cassisi
\altaffilmark{2}, Icko Iben, Jr. \altaffilmark{3},
and Amedeo Tornamb\'e \altaffilmark{2,4}}

\altaffiltext{1}{Dipartimento di Fisica, Universit\`a degli studi 
di Napoli ``Federico II'', Mostra d'Oltremare, Pad. 19, 80125 
Napoli; lpiersanti@astrte.te.astro.it}

\altaffiltext{2}{Osservatorio Astronomico di Teramo, Via M.Maggini 47,
64100 Teramo, Italy; cassisi@astrte.te.astro.it}

\altaffiltext{3}{Astronomy and Physics Departments, University of
Illinois, 1002 W. Green St., Urbana, IL 61801
icko@astro.uiuc.edu}

\altaffiltext{4}{Dipartimento di Fisica, Universit\`a de L'Aquila,
Via Vetoio, 67100 L'Aquila, Italy; tornambe@astrte.te.astro.it}


\begin{abstract}

Hydrogen-rich matter has been added to a carbon-oxygen white dwarf
of initial mass 0.516 \msun at the rates $10^{-8}$ and $2\times 10^{-8}$
\msun \yrm1, and results are compared with those for a white dwarf
of the same initial mass which accretes pure helium at the same rates.
For the chosen accretion rates, hydrogen burns in a series of recurrent
mild flashes and the ashes of hydrogen burning build up a helium
layer at the base of which a helium flash eventually occurs. In previous
studies involving accretion at higher rates and including initially more
massive white dwarfs, the diffusion of energy inward from the hydrogen
shell-flashing region contributes to the increase in the temperature at
the base of the helium layer, and the mass of the helium layer when the
helium flash begins is significantly smaller than in a comparison model
accreting pure helium; the helium shell flash is strong enough to cause
the model to expand beyond its Roche lobe, but not strong enough to
develop into a supernova explosion. In contrast, for the conditions
adopted here, the temperature at the base of the helium layer becomes
gradually independent of the deposition of energy by hydrogen shell flashes,
and the mass of the helium layer when the helium flash occurs is a
function only of the accretion rate, independent of the hydrogen content
of the accreted matter. Several thousand hydrogen shell flashes must be
followed before the helium flash takes place. Because of the
high degeneracy at the base of the helium layer, temperatures in the
flashing zone will rise without a corresponding increase in pressure,
nuclear burning will continue until nuclear statistical equilibrium
is achieved, and structural evolution will proceed hydrodynamically;
the model will become a supernova, but not of the classical type Ia
variety.

\end{abstract}
\noindent
{\em Subject headings:} stars: evolution --- stars: interiors --- 
supernovae: general --- stars: white dwarfs


\section{Introduction}

Recently, Cassisi, Iben, and Tornamb\'e (1998, hereinafter CIT) presented
results of an extensive set of numerical computations describing the long
term behavior of initially cool white dwarfs (WDs) accreting hydrogen-rich
matter. Initial masses of 0.516 \msun and 0.8 \msun were chosen and the
accretion rate was varied over the range $10^{-8}$ \msun \yrm1 to
$10^{-6}$ \msun \yrm1. For intermediate accretion rates in this range,
evolution was continued until a helium flash was initiated at the base
of the layer of helium built up by hydrogen burning. One of the main
results of the CIT experiments was a demonstration of how the thermal
structure of the helium layer is modified by the injection of energy due
to hydrogen burning compared with the thermal structure when pure helium
is accreted at the same rate. In general, for the same initial mass,
current mass, and accretion rate, the helium layer in the hydrogen-accreting
model is hotter (see Fig. 8 in CIT) than in the helium-accreting model, and,
when the helium flash begins, the total mass of the white dwarf accreting
hydrogen is smaller (with a correspondingly lower density in the helium
layer) than that of the white dwarf accreting pure helium. As a result,
the helium flash in the hydrogen-accreting model is much less powerful
than in the pure helium-accreting model\footnote{Note that, in the $3^{\rm d}$
line from the end of \S 1 in CIT, the words ``helium'' and ``hydrogen''
are mistakenly interchanged.}.

CIT also show that the envelopes of hydrogen-accreting white dwarfs
of low mass which suffer a violent, albeit probably not dynamical, helium
flash expand so much that, in the real world, most of the previously
accreted matter may be lost from the system because of Roche-lobe
overflow and common envelope activity.

In summary, in the CIT experiments, none of the white dwarfs accreting
hydrogen-rich matter are able to reach the Chandrasekhar mass (and
then explode their CO cores) or to experience a ``sub-Chandrasekhar''
detonation in their helium layers. However, CIT conjecture that models of
small mass which are accreting at rates which place them in the mild
hydrogen-flash regime, but approaching the strong hydrogen-flash regime,
may eventually experience a dynamical helium shell flash. Their conjecture
is based on the fact that, for a fixed (small) initial white dwarf mass,
the lower the accretion rate, the stronger is the final helium shell flash. 
By means of a simple extrapolation, CIT estimate a release of energy
during the helium flash which could eventually turn the burning into a
mildly dynamical event. The large number of hydrogen pulses required
to reach the helium flash stage in models accreting at the relevant small
rates persuaded CIT to postpone an explicit numerical investigation.

Over the past year, one relevant low-$\dot{M}$ calculation has been
carried to the helium shell flash stage and a second calculation has
reached the point where hydrogen-shell flashes are so strong as to
seriously slow the rate of progress towards a helium-burning thermonuclear
runaway. Even though not complete in the one case, both calculations 
establish that at the helium flash the mass of the helium layer and 
the violence of the event are much larger than expected on the basis
of extrapolation from CIT results. The reason for this is that, in the
present calculations, the helium layer becomes so massive prior to
the helium flash that it acts as a buffer, insulating the base of the
helium layer from the energy injected by hydrogen burning. Thus, the
mass of the He layer at He ignition is independent of the
composition of the accreted material.


\section{Results}

We have followed the very long term evolution of a white dwarf model
of mass 0.516 \msun accreting hydrogen at rates $10^{-8}$ and $2\times
10^{-8}$ \msun \yrm1. The initial white dwarf has a central carbon-oxygen
(CO) core of mass 0.48 \msun and an outer helium layer of mass 0.036 
\msunend. Details of the code, input physics, and numerical
assumptions can be found in CIT and references therein. For both
accretion rates, the accreting models undergo hydrogen shell
flashes which require both fine zoning and very small time steps over
a major portion of each pulse cycle. Typically, more than 2000 models
are required to follow in detail one single pulse cycle during which
the helium layer accretes about $10^{-4}$ \msun of fresh helium
produced by hydrogen burning.

Some characteristics of the model accreting at the rate $2\times
10^{-8}$ \msun \yrm1 are shown in Figure 1. There, the locations in mass
of the hydrogen-helium discontinuity ($M_{\rm H-shell}$), the point where
the temperature is at a maximum in the helium layer ($M_{\rm T-max}$), and 
the maximum in the helium-burning energy-generation rate
($M_{\rm He-shell}$) are shown as a function of the total mass
(i.e., time). During the early part of the evolution, $M_{\rm He-shell}$
is relatively close to $M_{\rm H-shell}$, as heat injected by hydrogen
shell flashes diffuses inward, maintaining high temperatures. However,
as temperatures and densities throughout the helium layer increase, the
peak in the helium-burning energy-generation rate shifts to the base
of the helium layer. This occurs quite suddenly at $M_{\rm tot} \sim
0.556$ \msunend.


Unfortunately, when the mass of the model reached
$\sim 0.645$ \msunend, in frustration at the real time consumed in
following hydrogen shell flashes, we increased the accretion rate just
enough that hydrogen began to burn in steady state. As may be deduced
from Figure 10 in CIT, the required accretion rate is $\sim 6 \times
10^{-8}$ \msun \yrm1. When model mass reached $\sim 0.66$ \msunend,
the accretion rate was returned to $2 \times 10^{-8}$ \msun \yrm1 and
hydrogen burning again proceeded by way of mild shell flashes. On
comparing the shape of the $M_{\rm T-max}$ curve before and after
the episode of steady-state burning, it is clear that the model never
completely recovered from the frustration-induced thermal pulse hiatus.
 
This failure to recover completely is demonstrated further in Figure 2
where the solid curve gives $\log T$ at the base of the helium layer as
a function of $\log \rho$ there. The dashed curve in Figure 2 gives
the same thing for the model which accretes pure helium at the rate
$2 \times 10^{-8}$ \msun \yrm1. The two curves do not begin at the
same point since quantities at the base of the helium layer were not
stored during the early evolution of the model accreting hydrogen-rich
matter. It is evident that, after the thermal pulse hiatus, the solid
temperature-density curve proceeds at a higher temperature level than
it did prior to the hiatus. Thus, the model retains a memory of the extra
energy injected by increasing, even briefly, the accretion rate by a
factor of 3. The dotted curve in Figure 2 is our conjecture for where
the $\log T$-$\log \rho$ curve would be had we not introduced the thermal
pulse hiatus. 


In the model accreting 
hydrogen-rich matter, the helium-burning thermonuclear runaway occurs
when model mass reaches $\sim 0.74$ \msunend, corresponding to a helium
layer of mass $\sim 0.26$ \msunend. The helium accreting model experiences
a thermonuclear runaway when model mass reaches $\sim 0.77$ \msunend,
corresponding to a helium layer of mass $\Delta M_{\rm He} \sim 0.29$
\msunend. Our conjecture (represented by the dotted curve in Fig. 2)
suggests that, had we not introduced the thermal pulse hiatus, the
characteristics at the base of the helium layer in the hydrogen-accreting
model would match those at the same point in the helium-accreting model
prior to the onset of the thermonuclear runaway. Thus, the total model
mass at the onset of the thermonuclear runaway is $\sim 0.77$ \msun in
both cases. The experiments with models accreting at the rate $10^{-8}$
\msun \yrm1 (see below) support our conjecture.

Extrapolating from the CIT results, we had expected that a helium-burning
thermonuclear runaway would occur when $\Delta M_{\rm He} \sim 0.12$
\msunend, less than half that given by the detailed computations. The
reason for this difference is clear. The larger $\Delta M_{\rm He}$ is,
the less sensitive (and lower) are temperatures in the deep interior of
the helium layer to the details of hydrogen burning above the helium layer.
The lower the accretion rate, the larger does $\Delta M_{\rm He}$ become
before the helium flash begins. Thus, the lower the accretion rate,
the closer does the pre-helium flash value of $\Delta M_{\rm He}$ for a
hydrogen-accreting model approach the pre-helium flash value of
$\Delta M_{\rm He}$ for its helium-accreting counterpart.

None of the models computed in the $\dot{M} = 2 \times 10^{-8}$ \msun
\yrm1 experiment enter the strong hydrogen-flash zone, therefore
it has been possible to follow the early development of the helium flash
up to the point that the outer edge of the convective layer formed
initially near the base of the helium layer reaches the base of the
hydrogen-rich envelope. Computations have been halted at this point,
because a realistic calculation requires one to follow the time dependent
diffusion of hydrogen into the helium convective zone until it reaches
temperatures at which hydrogen can ignite. One might anticipate that
hydrogen burning would force the development of a second, detached
convective layer and that the accretor in a real counterpart would
expand beyond its Roche lobe. 

However, a more dramatic fate is in store for the model. At the start
of the helium flash, the density ($\log \rho \sim 6.27$) and temperature
($\log T > 7.9$) at the base of the helium layer are such that electrons
are quite degenerate ($\epsilon_{\rm F}/kT > 20$), and one may expect that
most of the nuclear energy liberated in the flashing zone will initially
be converted into the thermal energy of ions and that, with temperatures
exceeding $10^9$ K, matter in the flashing zone will achieve nuclear
statistical equilibrium before expansion becomes important. The net
result will be a hydrodynamical event of supernova proportions (see the
discussion and references in Iben \& Tutukov 1991).

Additional hints as to the final outcome can be derived by comparing 
the physical properties of our model with those of Woosley \& Weaver (1994; 
hereinafter WW) at runaway. The WW experiments do not include any with
exactly the same initial mass and helium-accretion rate as we have used;
however, we can take advantage of the analytical study of the physical
properties of helium shell flashes provided by Fujimoto \& Sugimoto
(1982; hereinafter FS) to find a relevant WW model. The parameters of our
model in the $\Delta{M_{\rm He,pk}}/M_{\odot} - M_{\rm WD}$ plane are
slightly outside of the region for which the FS assumptions are strictly
valid, but they are close enough that the FS results can be extrapolated
with some confidence to estimate
that the maximum energy-generation rate of our model is comparable to 
that of the WW ``model 1'': $ M_{\rm WD,0} =0.6 M_{\odot},
\dot{M}= 2.5 \times 10^{-8}$ \msun \yrm1. This model produces an
outgoing detonation in the helium layer and an inward moving sonic wave
which eventually focuses at the center of the white dwarf causing
an induced thermonuclear runaway. All in all, $\sim 0.9 \times 10^{51}$
ergs will be delivered in the event and the star will be completely
disrupted. 

The hydrogen accretion experiment with $\dot{M}=10^{-8}$ \msun \yrm1
has been followed, with no change in the accretion rate (i.e., no
artificially introduced thermal pulse hiatus) until it has become clear
that, long before the onset of the helium flash, the thermal structure
near the base of the helium layer approaches that near the base of the
helium layer in the companion model which accretes pure helium. The
results which demonstrate this are shown in the temperature-density
plane of Figure 3.


The calculation of the hydrogen-accreting model has been terminated
because the hydrogen shell flashes have become so strong as to make
following the evolution further impractical. At termination,
$\Delta M_{\rm He} \sim 0.32$ \msun and $M_{\rm tot} \sim 0.80$ \msunend.

Despite the termination, one can speculate reasonably confidently about
further evolution. The hydrogen flashes are strong enough that the
expansion of the hydrogen-rich envelope in a real analog extends well
beyond the Roche lobe, with concomitant loss of matter of the
sort experienced by classical novae. Thus, the real analog may never
reach the helium-flashing stage. If it does reach the helium-flashing stage,
one may use the properties of the helium-accreting model to predict that
a supernova explosion terminates evolution. At the start of the helium
flash, $\Delta M_{\rm He} \sim 0.50$ \msun and $M_{\rm tot} \sim 0.98$
\msunend; $\log \rho \sim 6.9$, $\log T \sim 7.8$, and $\epsilon_{\rm F}/kT
> 200$ at the base of the helium layer. These numbers imply that the
supernova explosion terminating the life of the real analog, if it can
reach this stage, will be even more violent than the one predicted for
the model accreting hydrogen-rich matter at twice the rate.
    
In conclusion, we have shown that, for a restricted range of
hydrogen-accretion rates and initial CO white dwarf masses, the
outcome is a ``sub-Chandrasekhar'' supernova. In the absence of a
hydrogen-rich companion, the real analog of the exploding white
dwarf model would probably appear as a type I supernova (see, e.g., WW).
That is, the hydrogen mass present at the top of the exploding dwarf
($\sim 10^{-4}$ \msunend) is probably too small to produce detectable
hydrogen lines in the spectrum. However, the nucleosynthetic products of
a sub-Chandrasekhar explosion are such that the spectrum will depart
considerably from that of a classical type Ia supernova (H\"oflich \& 
Khokhlov 1996, Nugent et al. 1997, several papers in Truran \& Niemeyer
1999). In the real analog, contamination of the supernova ejecta by
hydrogen from the hydrogen-rich companion may even move the supernova
out of the type I category entirely. The range of parameters
for which an explosive outcome of the sort described here will occur
suggests a variation in explosion energy of slightly less than a factor
of 2. An estimate of what fraction of observed supernovae are of the sort
discussed here requires a full fledged population synthesis study;
such an estimate is beyond the scope of this paper.

\section{Influence of NCO Reactions on Flash Properties}

The referee has noted that, in the computations just described, the NCO
reactions $^{14}$N(e$^{-},\nu)^{14}$C($\alpha,\gamma)^{18}$O were not
taken into account. There are several reasons for this omission. First,
our starting white dwarf model is the end product of an evolutionary
calculation with mass loss and, during the conversion of the helium core
into a CO core, nearly all of the $^{14}$N initially present in the core
is destroyed via the $^{14}$N$(\alpha,\gamma)^{18}$O$(\alpha,\gamma)^{22}$Ne
reactions, leaving only a very thin layer of unburned helium near the top.
In fact, in the initial white dwarf model, there is no $^{14}N$ at all
below the point where the helium abundance is $\sim 0.7$, and, in our
accretion models, helium ignition occurs much below this point. Secondly,
a comparison between the results of Taam (1980) and of Woosley, Taam,
\& Weaver (1986), who do not take the NCO reactions into account, and
the results of WW, who do take them into account, shows that the inclusion
of the NCO reactions lowers the critical mass of the accreted layer at
the start of the triple-$\alpha$ runaway but does not prevent an explosive
outcome, as the referee suggests might be the case. The latter three
computations begin with a pure CO white dwarf which does not have the
surface helium layer (in most of which there is no $^{14}$N) which white
dwarfs derived by evolutionary calculations possess. A comparison between
results of the WW computations and those of Limongi \& Tornamb\`e (1991),
who used initial models derived by evolution, suggests that the influence
of the NCO reactions on inducing the triple-$\alpha$ runaway may be quite
small when evolutionary models are used, due to the lack of $^{14}N$ nuclei
over the inner part of the helium layer capping the CO core in the initial
model.

Nevertheless, following the referee's suggestion, we have analyzed
explicitly how inclusion of the NCO reactions alters our present results.
We employ the Hashimoto et al. (1986) rate for the
$^{14}$N(e$^{-},\nu)^{14}$C reaction and the Caughlan, Fowler \& Zimmermann
(1988) rate for the $^{14}$C$(\alpha,\gamma)^{18}O$ reaction. The abundance
of $^{14}$N in accreted material is taken as solar. As expected, NCO burning
begins far from the base of the helium layer, inside the region where
$^{14}N$ nuclei are present at densities such that the threshold for
electron capture on $^{14}$N is exceeded. 
As NCO burning becomes active, $3\alpha$ reactions are already occurring
at a non-negligible rate at the base of the helium layer. As accretion
continues, the NCO-burning shell moves outwards in mass in consequence
of the general increase in density in the helium shell and of the outward
(in mass) motion of the base of the $^{14}$N$\rightarrow ^{14}$C
transition zone.

While the peculiar behaviour of NCO burning is interesting in its own right
(we will address this topic in some detail in a forthcoming paper), its
actual effect on the outcome of our helium-accreting CO white dwarfs is
rather small. As a matter of fact, in the $\dot{M}_{\rm He}=10^{-8}$
\msun \yrm1 experiment, the differences are negligible (less than a
2\% change in both the mass of the accreted layer and in the ignition
density when the thermonuclear runaway begins. Interestingly, the effect
is slightly more important in the model accreting at the rate
$\dot{M}_{He}=2 \times 10^{-8}$ \msun \yrm1, but, even so, the differences
in ignition density and in the critical mass of the helium layer at
the $3\alpha$ runaway are less than 10\%. By interpolation, we estimate
that, for an accretion rate of $\dot{M}_{He}=1.58 \times 10^{-8}$ \msun
\yrm1, triple-$\alpha$ runaway conditions will be essentially identical
to those found in the $\dot{M}_{He}=2 \times 10^{-8}$ \msun \yrm1 model
without the NCO reactions. Our previous arguments for a dynamic event are
thus still valid.

It is easy to understand why, the larger the accretion rate, the more
effective are the NCO reactions in reducing the critical mass of the
helium layer when the $3\alpha$ runaway begins, relative to the
case when the NCO reactions are neglected. With or without the NCO
reactions, increasing the accretion rate leads to more rapid heating and
a decrease in the critical mass for helium ignition. Thus, with the NCO
reactions included, the larger the accretion rate, the closer to the base
of the helium layer does the NCO-burning shell form. The closer the
NCO-burning shell is to the base of the helium layer, the more rapidly
does heat from this shell diffuse to the base of the helium layer and the
sooner are conditions reached for the triple-$\alpha$ runaway to begin.
A similar phenomenon occurs in models accreting hydrogen-rich matter at
a high rate (CIT); the larger the accretion rate, the greater is the
influence of heat diffusing from the hydrogen shell flashing region in
inducing helium-burning flashes.

The most important result of the experiments described in this section
is that explosive outcomes are produced in the range of $\dot{M}$ examined,
even when NCO reactions are taken into account, thus confirming previous
results by WW.

%
%
%

\newpage


\figcaption{Several characteristics of a model white dwarf accreting
hydrogen-rich material at the rate $\dot{M} = 2 \times 10^{-8}$ \msun
\yrm1 are shown as a function of the total mass (i.e., of the time). 
Characteristics are the locations in mass of the hydrogen-helium
discontinuity ($M_{\rm H-shell}$), the point where the temperature
is at a maximum in the helium layer ($M_{\rm T-max}$), and the maximum
in the helium-burning energy-generation rate ($M_{\rm He-shell}$).
\label{fig1}}


\figcaption{The evolution in temperature $T$ and density $\rho$ at the
base of the helium layer in model white dwarfs accreting hydrogen-rich
matter (solid line) and pure helium (dashed line) at the rate $\dot{M} =
2 \times 10^{-8}$ \msun \yrm1. The dotted curve is an estimate of
behavior had the hydrogen flashes not been temporarily suppressed by
increasing the accretion rate by a factor of three as total mass is
increased from $\sim 0.645$ \msun to $\sim 0.66$ \msunend.
\label{fig2}}


\figcaption{Same as Fig. 1 when $\dot{M} = 2 \times 10^{-8}$ \msun \yrm1,
except that the accretion rate has been kept strictly constant for the
entire evolution of the hydrogen-accreting model.
\label{fig3}}

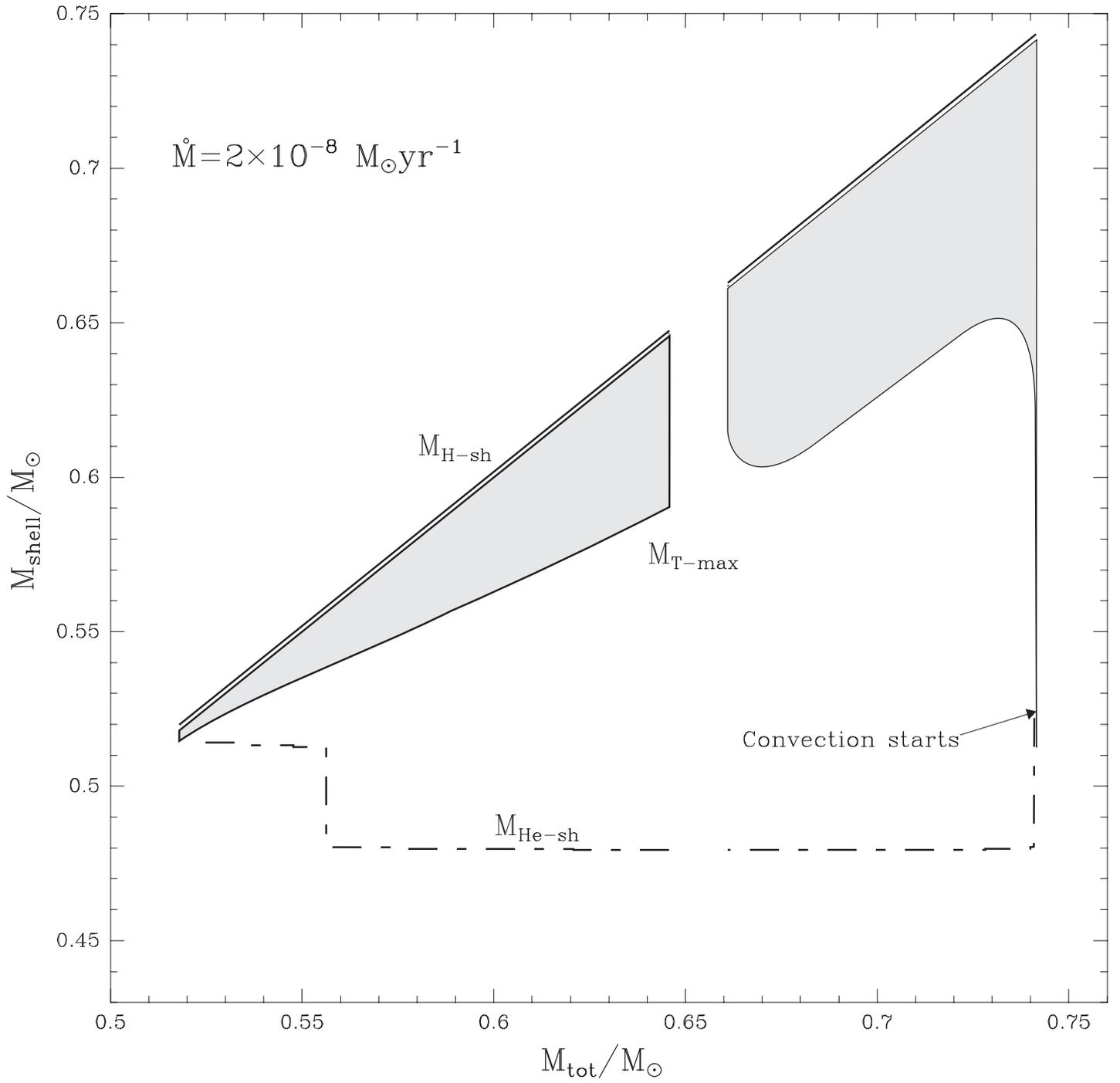
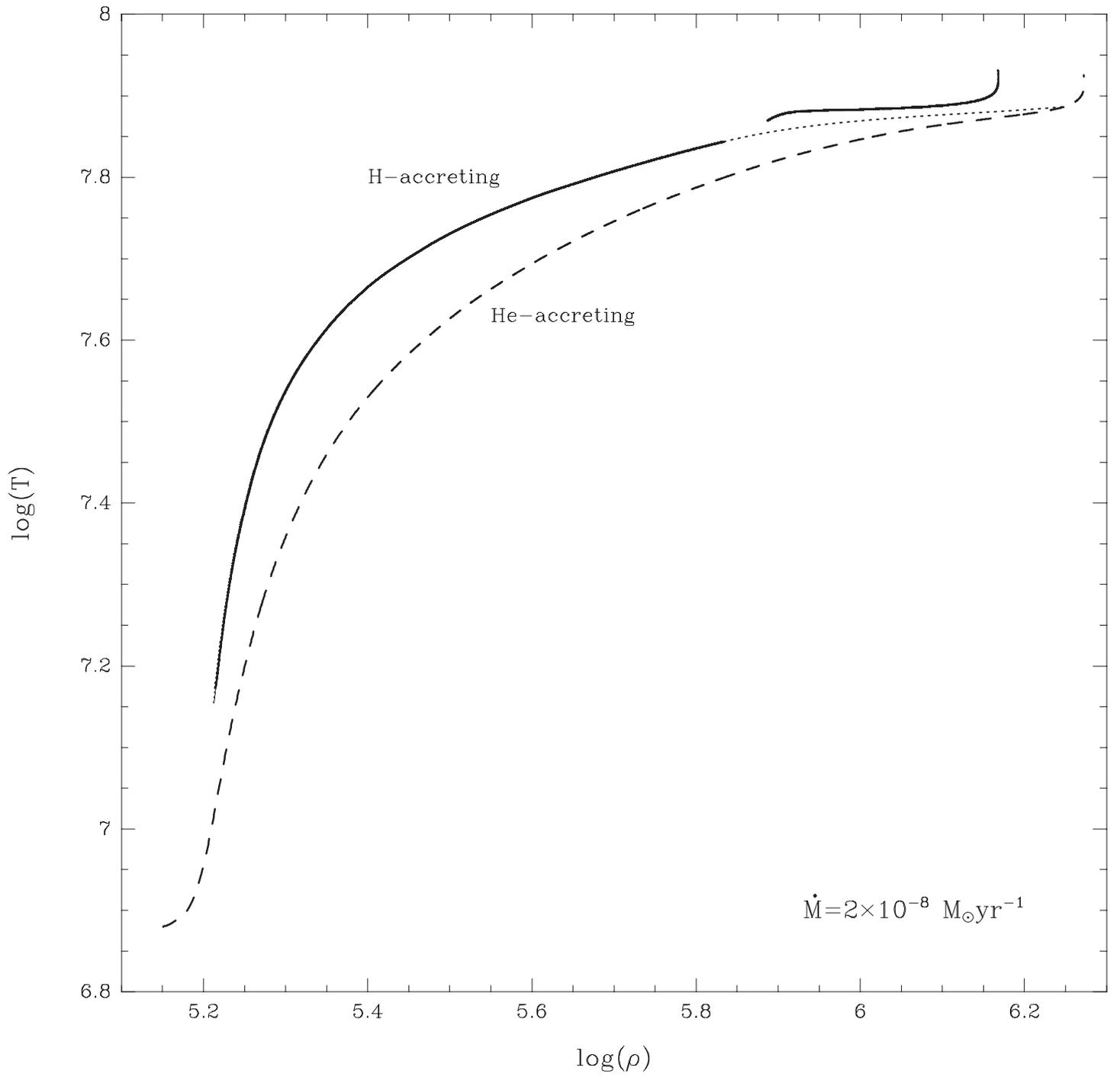
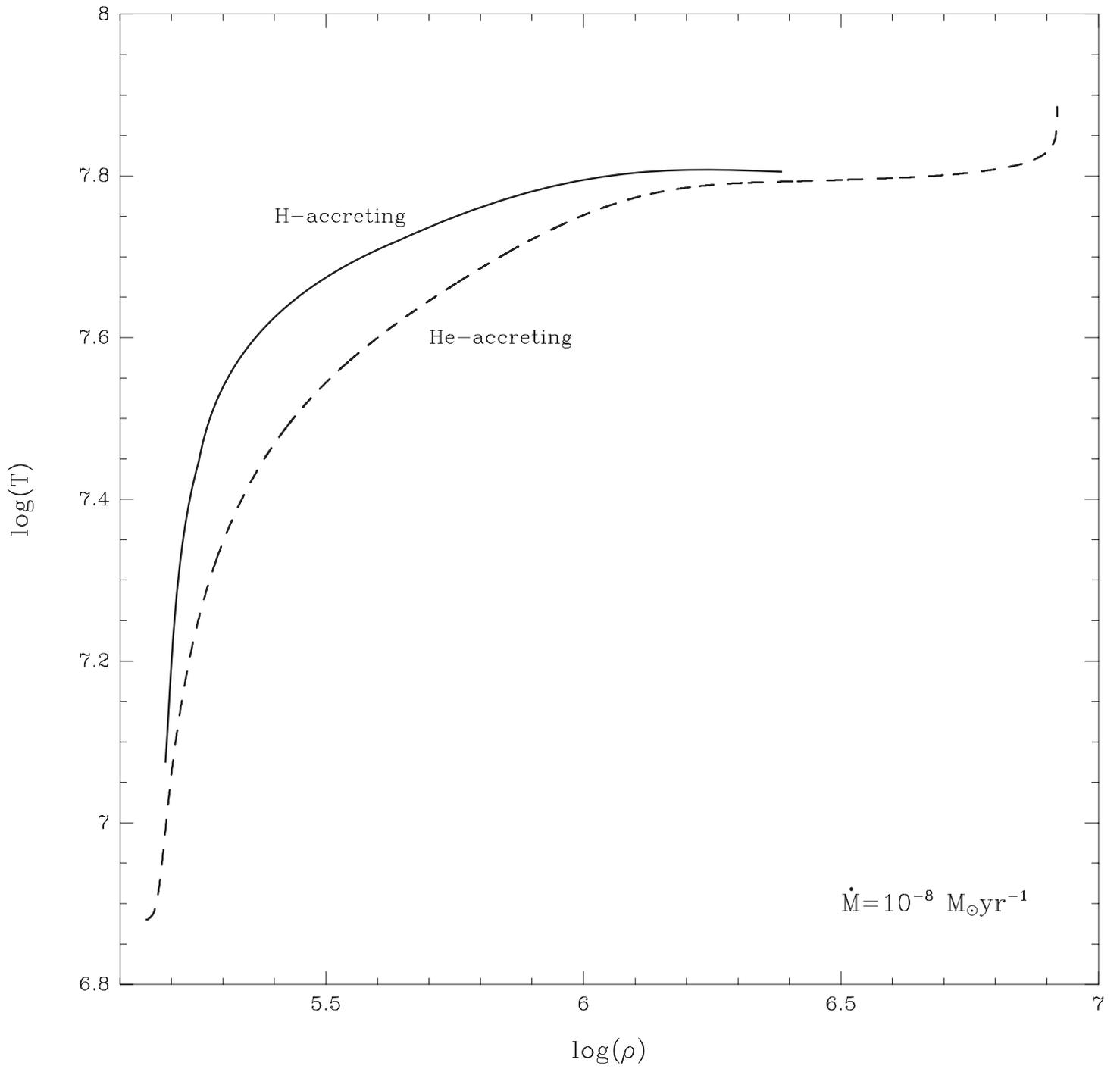
\end{document}